\documentclass{raa}
\usepackage{graphicx,times}             %for PS/EPS graphics inclusion, new
\input{epsf.sty}                        %for PS/EPS graphics inclusion, old
\input{psfig.sty}                       %for PS/EPS graphics inclusion, old

\begin{document}

   \title{The properties of a large volume-limited sample of face-on low surface brightness disk galaxies
%\,$^*$
%\footnotetext{$*$ Supported by the National Natural Science Foundation of China.}
}
%   \subtitle{I. Place Your Subtitle Here}

   \volnopage{Vol.0 (200x) No.0, 000--000}      %%preserved for Editor. DOn't remove!
   \setcounter{page}{1}          %%starting page, preserved for Editor. DOn't remove!

   \author{G.-H. Zhong
      \inst{1,2,3,4}
   \and Y.-C. Liang
      \inst{1,2}
   \and F.-S. Liu
      \inst{5}
   \and F. Hammer
      \inst{4}
   \and K. Disseau
      \inst{4,6}  
   \and   L.-C. Deng  
      \inst{1,2}  
   }
%% Here is an example of three authors come from different institutes.
%% For single author or all the authors from an institute, use "\inst{}" only

   \institute {National Astronomical Observatories, Chinese Academy of Sciences, Beijing 100012, China; {\it ghzhong@nao.cas.cn}\\
%% Please give the E-mail address of the author, to whom future correspondence and
%% offprint requests will be sent.
        \and
             Key Laboratory of Optical Astronomy, National Astronomical Observatories, Chinese Academy of Sciences, Beijing 100012, China \\
        \and
     Graduate University of Chinese Academy of Sciences, Beijing 100049, China \\
        \and
    GEPI, Observatoire de Paris-Meudon, 92195 Meudon, France \\
        \and
    College of Physical Science and Technology, Shenyang Normal University, Shenyang 110034, China \\
        \and
    	Univ Paris Diderot, Sorbonne Paris Cité, GEPI, UMR8111 CNRS, Observatoire de Paris, F-92190 Meudon, France
   }

   \date{Received~~2012 month day; accepted~~2012~~month day}

\abstract{We select a large volume-limited sample of low surface brightness
galaxies (LSBGs, 2,021) to investigate their statistical properties 
and their differences from high surface brightness galaxies (HSBGs, 
3,639) in details. The distributions of stellar masses of LSBGs and 
HSBGs are nearly the same and they have the same median values.  
Thus this volume-limited sample have good completeness and further 
remove the effect of stellar masses on their other properties when 
we compare LSBGs and HSBGs. We found that LSBGs tend to have lower 
stellar metallicities, and lower effect dust attenuations indicating 
that they have lower dust, than HSBGs. The LSBGs have relatively 
higher stellar mass-to-light ratios, higher gas fraction, lower star 
forming rates (SFRs), and lower specific SFRs than HSBGs. Moreover, 
with the decreasing surface brightness, gas fraction increase, while 
the SFRs and specific SFRs decrease rapidly for the sample galaxies. 
This could mean that the star formation histories between LSBGs and 
HSBGs are different, HSBGs may have stronger star forming activities 
than LSBGs.
\keywords{galaxies: fundamental parameters, galaxies: general, galaxies: statistics, galaxies: stellar content}
}

   \authorrunning{G.-H. Zhong, et al.}            %author_head in even pages
   \titlerunning{The properties of a large volume-limited sample of face-on LSBGs}  % title_head in odd pages

   \maketitle
%% The author head (on even pages) and the title head (on odd pages) will be
%% automatically extracted from \author{} and \title{}. Whenever the title is too long,
%% you will be asked to supply a shorter one by inserting either \authorrunning{} or
%% \titlerunning{} before \maketitle. Anyway, you can specify your own heads.
%%
%%
%% Note: In the following text body of your manuscript, please note several differences from
%%       other major journals:
%% (1) \subsection{Please Capitalize the First Letter of Each Notional Word in Subsection Title}
%% (2) Please Capitalize the First Letter of Each Notional Word in all tables' captions

%
%________________________________________________ sections below
%
\section{Introduction}
\label{introduction}

Galaxies with surface brightness fainter than $\mu_0(B)
=$ 21.65$\pm$0.3 mag arcsec$^{-2}$ are well known as low surface
brightness galaxies (LSBGs, Freeman 1970). Yet, owing to their
faintness compared with the night sky, they are hard to find.
Hence their contribution to the local galaxy population has been
underestimated for a long time. However it is found that LSBGs
could represent a significant fraction of galaxy number density
in the universe (O'Neil \& Bothun 2000; Minchin et al. 2004;
Trachternach et al. 2006) and may comprise up to half of the
local galaxy population (McGaugh et al. 1995).

During the last four decades, LSBGs are studied widely both in
observations (Impey et al. 1996; O'Neil et al. 1997; Trachternach
et al. 2006; Haberzettl et al. 2007a; Pizzella et al. 2008; Ulmer 
et al. 2011; Morelli et al. 2012) and theoretical work (Dalcanton et al. 
1997; Alard 2011). Especially, thanks to the modern digital sky surveys,
a wealth of observational data, with high quality, is undoubtedly
important to allow us to study the photometric and/or spectroscopic
properties of LSBGs with large samples and even in multi-wavelengths.
For example, (1) the statistical properties (Zhong et al. 2008, 2010),
the metallicities (Liang et al. 2010), the environment (Rosenbaum et
al. 2004, 2009; Galaz et al. 2011), and the stellar red holes (Bergvall
et al. 2010) of LSBGs from Sloan Digital Sky Survey (SDSS); (2)
the stellar populations (Zhong et al. 2008) and HI observation (Monnier
Ragaigne et al. 2003a, 2003b) of LSBGs from Two Micron All Sky Survey
(2MASS); (3) Star Formation Efficiency (Boissier et al. 2008) from
Galaxy Evolution Explorer (GALEX); (4) infrared properties (Hinz et al.
2007) from Spitzer Space Telescope. In addition, the multi-wavelength
SEDs are also investigated (Gao et al. 2010).

From the studies above, LSBGs, generally, are found to have lower metallicities
(McGaugh 1994; Galaz et al. 2006; Haberzettl et al. 2007b; Liang et
al. 2010), lower surface densities that could explain their slow evolution
(Mo et al. 1994; Gerritsen \& de Blok 1999; van den Hoek et al. 2000),
lower fraction of AGN (Impey et al. 2001; Mei et al. 2009; Liang et al.
2010; Galaz et al. 2011), lower star formation rate (SFRs, van den Hoek
et al. 2000), higher $M_*/L$ ratio (Sprayberry et al. 1995), higher
gas fraction, and they are located in lower densities regions (Mo et al. 1994;
Rosenbaum et al. 2004, 2009; Galaz et al. 2011) than what is typically found
in HSBGs.

Despite these impressive progresses, there are still several challenges
about LSBGs, such as many aspects of their formation and evolution.
Moreover, the studies of $M_*/L$, SFRs or gas contents of LSBGs are based on
very small samples, less than two hundred LSBGs (e.g. McGaugh \& de Blok 1997;
Burkholder et al. 2001). In particular, a magnitude-limited survey will
be affected by selection effects, which result from the inability of the
survey to detect fainter galaxies at larger redshifts (e.g. Impey et al. 1996,
Zhong et al. 2008). One way to avoid these effects is through using
a volume-limited sample, in which a maximum redshift and minimum absolute
magnitude are chosen so that a complete sample is obtained in this
redshift and magnitude ranges. Some properties could be different
between the volume-limited sample and the magnitude-limited sample. For
example, in Zhong et al. (2008), the relation between disk scale-length
and $\mu_0(B)$ are different between total sample and the volume-limited
sample. This difference is caused by the selection of volume-limited
sample from the magnitude-limited sample. When selecting volume-limited
LSBGs at $z <$ 0.1, we exclude two parts: on the one hand, LSBGs with
relatively lower redshift (smaller distance) and lower luminosity (see  
bottom-left of Fig. \ref{fig.mr}) are excluded, which should have smaller 
disk scale-length (see Fig. 4 and Fig. 5 of Zhong et al. 2008) as well as 
fainter surface brightness (see Fig. 3 of Zhong et al. 2008); on the other
hand, LSBGs with relatively higher redshift (larger distance) and higher 
luminosity (see top-right of Fig. \ref{fig.mr}) are excluded, which should 
have larger disk scale-length (see Fig. 4 and Fig. 5 of Zhong et al. 2008) 
as well as brighter surface brightness (see Fig. 3 of Zhong et al. 2008). 
Excluding LSBGs with smaller disk scale-length at fainter surface brightness
end and larger disk scale-length at the brighter surface brightness end 
from Fig. 7 of Zhong et al. (2008) results in the relation between disk 
scale-length and surface brightness of their Fig 13g, the more obvious 
correlation between log$h$ and $\mu_0(B)$. Furthermore, in Galaz 
et al. (2011), they point out that the strong dependence of the absolute 
magnitude versus redshift in the SDSS spectroscopic catalog (a magnitude-limited 
catalog), and the trend of the absolute magnitude on the galaxy size (see 
Fig. 1 of Galaz et al. 2011 for more details).

 In this work, with the advent of the large sky survey of the SDSS, it 
is now possible to dramatically extend the studies of SFRs, $M_*/L$, and gas 
contents of LSBGs in sample size. Moreover, this large amount of high quality 
data will be undoubtedly important to allow us to study the properties
of those galaxies more carefully. Furthermore, in order to avoid the bias
of magnitude-limited samples, we constrain on a large volume-limited sample
of LSBGs (more than two thousands) to study their statistical properties in 
detail, and then compare them with HSBGs. The volume-limited samples could
be a fair comparison between LSBGs and HSBGs, and they improve the completeness 
of the samples.

This is one of our series work (Zhong et al. 2008, 2010; Liang et al.
2010; Gao et al. 2010; Chen et al. in preparation) to study the properties
of a large sample of LSBGs and compare their differences with HSBGs. This paper
is organized as follows. We describe our sample selections in Sect. 2
and present the results in Sect. 3. In Sect. 4 we discuss our results.
In Sect. 5 we summarize this work.
Throughout the paper, a cosmological model with $H_0$ = 70 km s$^{-1}$
Mpc$^{-1}$, $\Omega_M$ = 0.3 and $\Omega_\lambda$ = 0.7 is adopted.

\section{Samples}
\label{sample}

The data analyzed in this study are drawn from the SDSS, which is
an imaging and spectroscopic survey that will obtain photometry of
one-quarter of the sky and spectra of nearly one million objects.
The imaging data were collected in $u,~g,~r,~i,~z$ band mounted on
the SDSS 2.5 m telescope at Apache Point Observatory. The spectra
are flux- and wavelength-calibrated, with 4,096 pixels from 3,800 to
9,200 \AA~ at $R \sim$ 1,800. The details of the survey strategy and
overview of the data pipelines and products can be found in York
et al. (2000) and Stoughton et al. (2002). In this paper, for the
sample galaxies, the stellar masses, mass-to-light ratios, and effective
dust attenuations are from Kauffmann et al. (2003); the SFRs and
specific SFRs are from Brinchmann et al. (2004); the stellar metallicities
 are from Gallazzi et al. (2005). All of those data can be
found in the MPA/JHU website\footnote{http://www.mpa-garching.mpg.de/SDSS/DR4}.

The same as Zhong et al. (2008), the sample is selected from the
main galaxy sample (MGS, Strauss et al. 2002) of SDSS Data Release
Four\footnote{http://www.sdss.org/DR4} (DR4, Adelman-McCarthy et al.
2006). We prefer to use the DR4 because some of the parameters
we use here have just been updated to DR4 by the MPA/JHU groups, e.g.
stellar metallicities from Gallazzi et al. (2005).
The detailed criteria of sample selection can be found in Zhong et
al. (2008). We describe it briefly as follows.

\begin{enumerate}

\item $fracDev_r <$ 0.25, the parameter $fracDev_r$ indicates the fraction
of luminosity contributed by the de Vaucouleurs profile relative to
exponential profile in the $r$-band. This is to select disk dominated
galaxies which can minimize the effect of bulge light on the disk galaxies.

\item $b/a$ $>$ 0.75 (corresponding to the inclination $i<$ 41.41 degree),
this is to select nearly face-on galaxies, which can minimize the extinction.
$a$ and $b$ are the semi-major and semi-minor axes of the fitted exponential
disk respectively.

\item $M_B$ $<$ -18, this is to exclude a few dwarf galaxies contained in the
sample.

\item We select objects with $\mu_0(B) \ge$ 22.0 mag arcsec$^{-2}$ as
LSBGs and $\mu_0(B) <$ 22.0 mag arcsec$^{-2}$ as HSBGs. The $\mu_0(B)$ are
calculated following the method of Zhong et al. (2008, their Eq. 6). After
this step, we get 12,282 LSBGs and 18,051 HSBGs.

\item In order to avoid the bias introduced from the differences in both
the redshift and absolute magnitude distributions for the samples, we extract
volume-limited samples from the $M_r-z$ plane by considering $z<0.1$ and
galaxies brighter than the corresponding $M_r$. After this step, we get
3,313 LSBGs and 4,722 HSBGs. \\
The relationship between $r$ band absolute magnitude and redshift for LSBGs
is showed in Fig.~\ref{fig.mr}. The vertical line refers to $z =$ 0.1, while
the horizontal line is the corresponding $M_r$. The galaxies located in the
top-left region within the solid lines are selected as our volume-limited
sample.

\begin{figure}
\centering
\includegraphics[width=7.0cm]{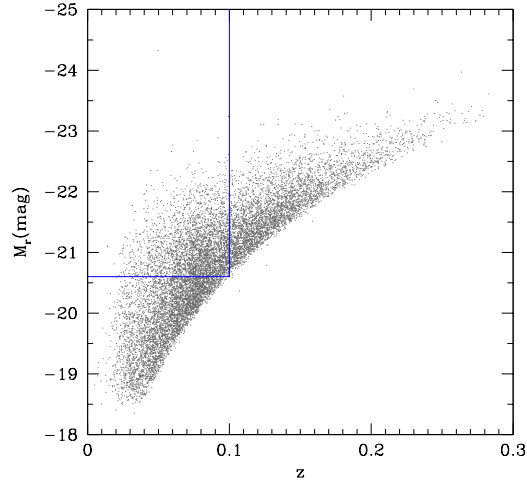}
\caption{The relationship between $r$ band absolute magnitude and redshift
for LSBGs. The vertical line refers $z =$ 0.1 and the horizontal line is
the corresponding $M_r$. The top-left region with solid lines is selected
as volume-limited sample.}
\label{fig.mr}
\end{figure}

\item In this work, in order to obtain the values of the property parameters,
we just select galaxies with all measurements of stellar masses, mass-to-light
ratios, effective dust attenuations, SFRs and specific SFRs, and stellar
metallicities. At last, we get the volume-limited samples
of 2,021 LSBGs and 3,639 HSBGs following the volume-limited sample selections
of Zhong et al. (2008).

\end{enumerate}

All the magnitudes, we quote above, are $K$-corrected and Galactic reddening
corrected (Blanton et al. 2005).

\section{Results}
\label{result}

Using large volume-limited samples of LSBGs and HSBGs, we present the
properties of LSBGs (e.g. stellar metallicities, effective dust
attenuations, mass-to-light ratios, and SFRs) and compare them with HSBGs.
The stellar metallicities are from Gallazzi et al.
(2005), in which they determine the stellar metallicities by a
spectroscopic method that used a few line strength indices to measure the
stellar population properties of galaxies. The stellar masses, mass-to-light
ratios, and effective dust attenuations in $z$ band are from Kauffmann et al. (2003),
all of which are generated by using a large library of Monte Carlo realizations
of different star formation histories, including starbursts of varying
strength and a range of metallicities. The SFRs are from Brinchmann et al.
(2004), who build up a picture of the nature of star-forming galaxies at
$z <$ 0.2 by comparing physical information extracted from the emission
lines (e.g. H$\alpha$) with continuum properties, and develop a method for
aperture correction using resolved imaging. This method takes out essentially
all aperture bias in their SFRs estimates, allowing an accurate estimates
of the total SFRs in galaxies. One may wish to refer to the papers of Gallazzi
et al. (2005), Kauffmann et al. (2003), and Brinchmann et al. (2004) about
the properties of galaxies in SDSS.

\subsection{Distributions of stellar masses, stellar metallicities, 
effective dust attenuations and stellar mass-to-light ratios}
\label{distrib}

\begin{figure*}
\centering
\includegraphics[width=7.0cm]{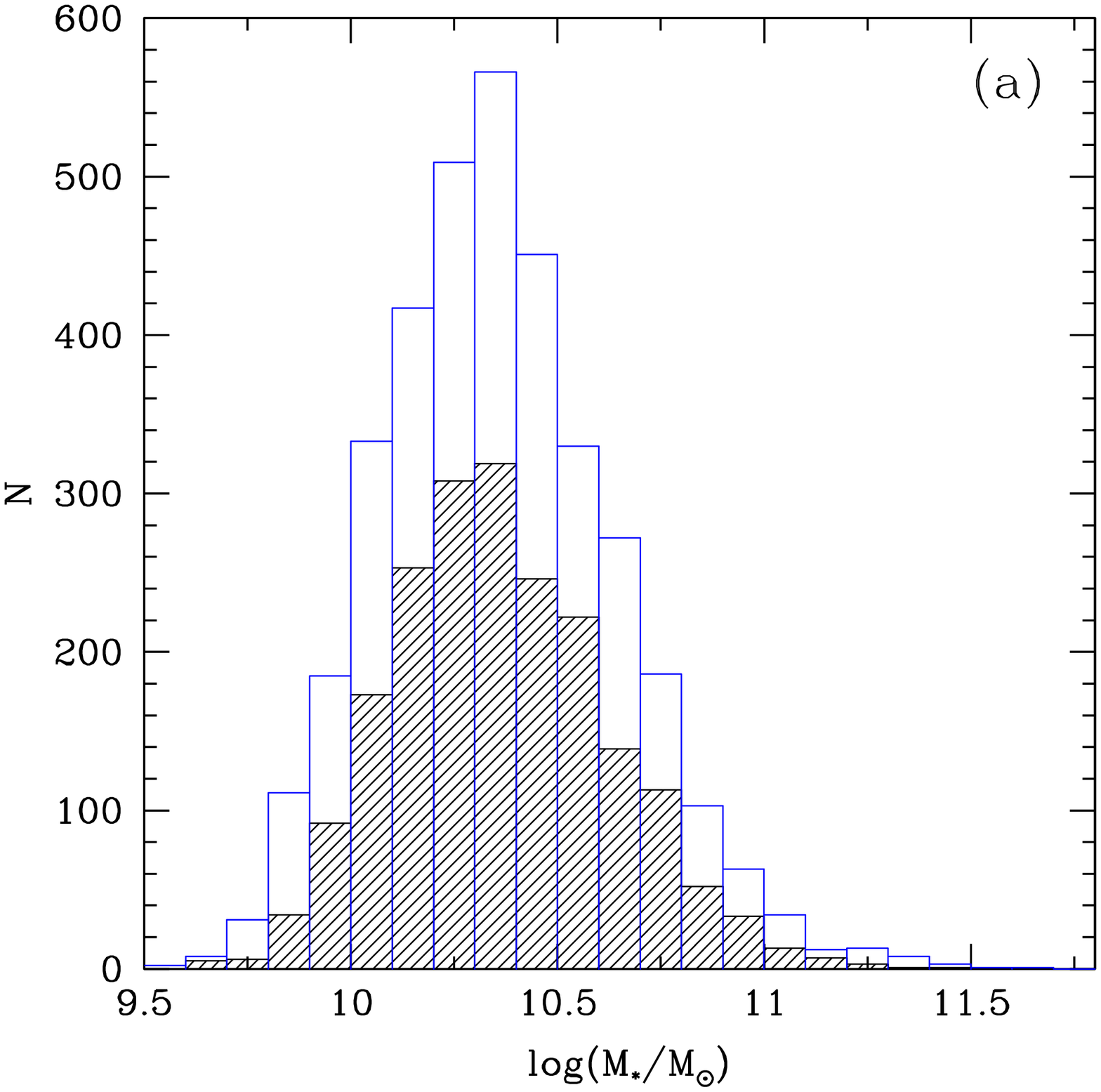}
\includegraphics[width=7.0cm]{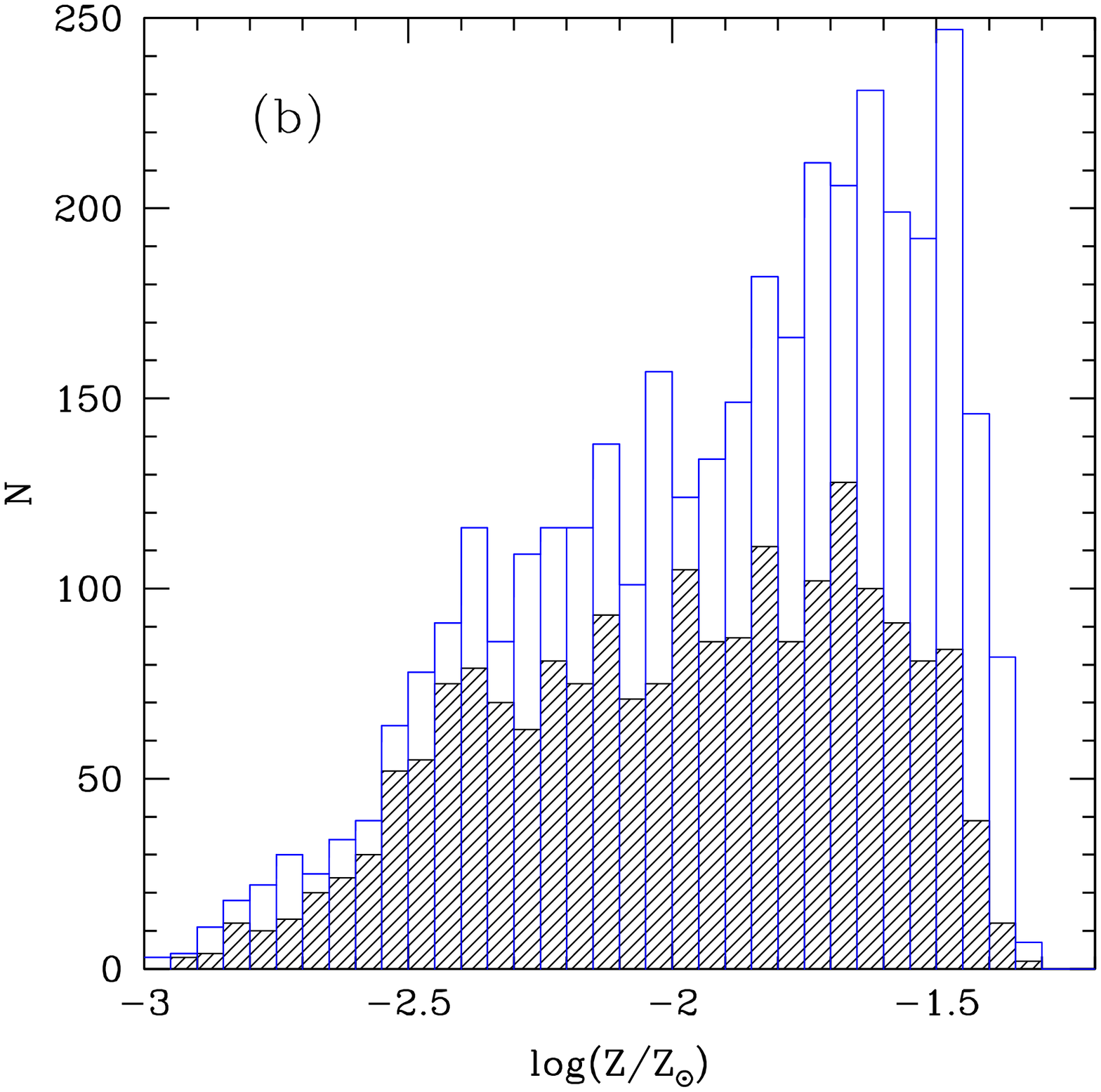}
\includegraphics[width=7.0cm]{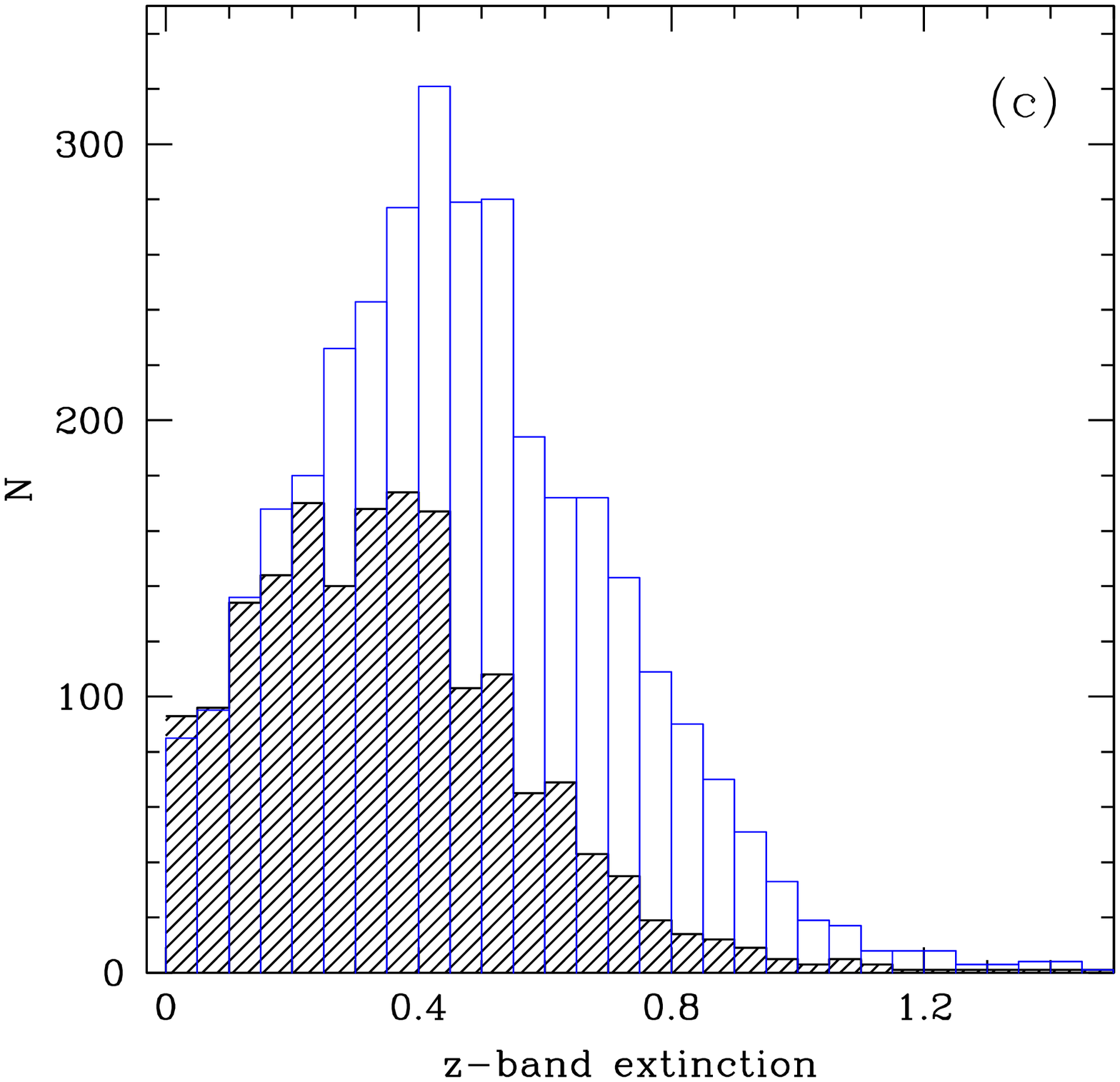}
\includegraphics[width=7.0cm]{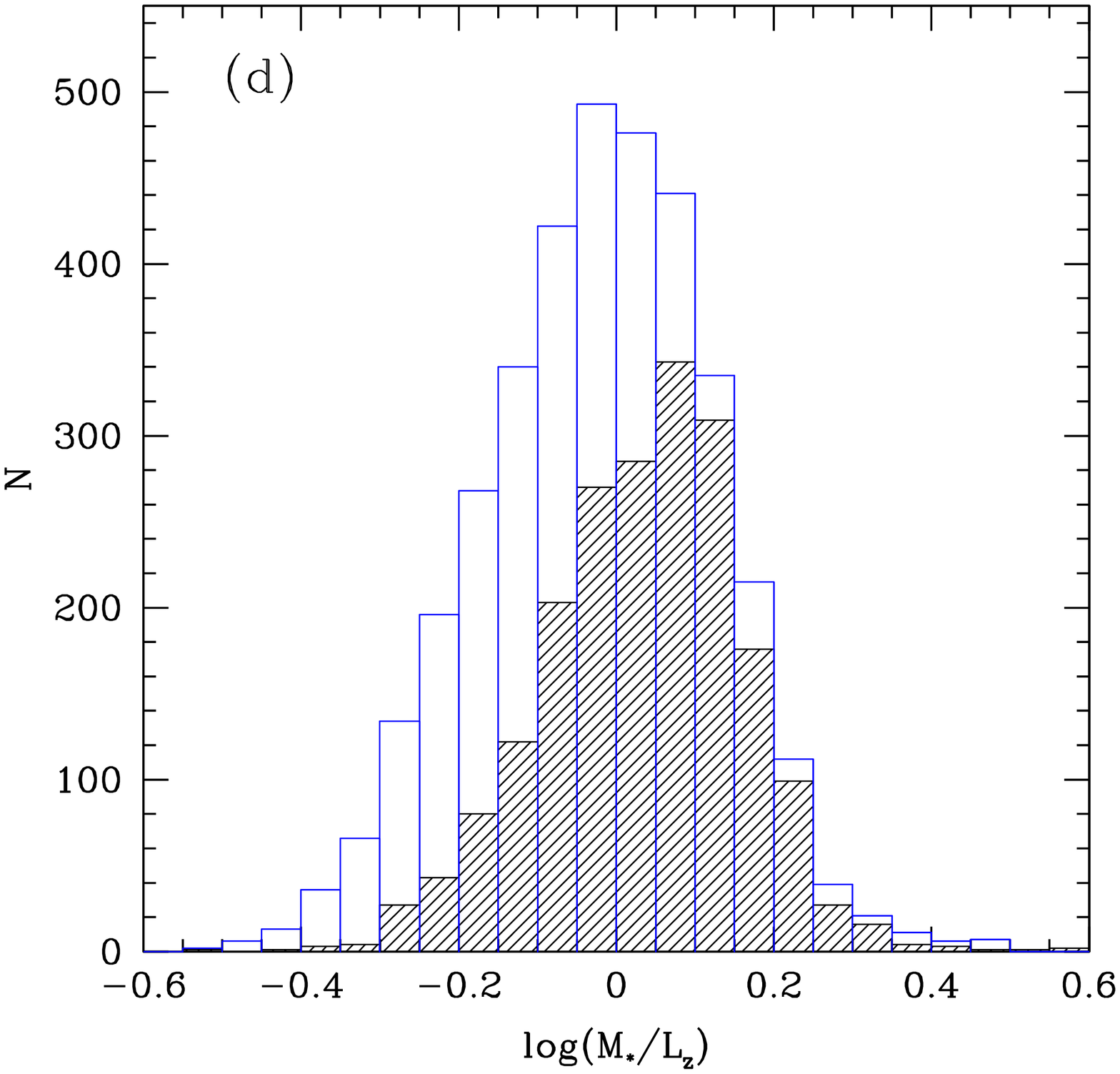}
\caption{Histograms distributions of some parameters for LSBGs (shadowed
regions) and HSBGs (white regions): (a). the distributions of stellar masses
(log($M_*/M_\odot$)), (b). the stellar metallicities (log($Z/Z_\odot$)), and (c).
the effective dust attenuations in $z$ band, (d) the histogram distributions 
of dust-corrected mass-to-light ratios in $z$ band.}
\label{fig.his}
\end{figure*}

Fig. \ref{fig.his} shows the histograms distributions of some
parameters for LSBGs and HSBGs. They are the distributions of
the stellar masses (Fig.~\ref{fig.his}a), the stellar metallicities
(Fig.~\ref{fig.his}b), effective dust attenuations in $z$ band
(Fig.~\ref{fig.his}c), and dust-corrected mass-to-light 
ratios in $z$ band (Fig.~\ref{fig.his}d), respectively. The 
shadowed regions are for LSBGs, while the white regions are for 
HSBGs.

From the distributions of stellar masses in Fig. \ref{fig.his}a,
we can see that LSBGs and HSBGs span nearly the same range and
are with very similar distributions. Kolmogorov-Smirnov tests
show that the different mean of distributions between LSBGs and
HSBGs is only 6.6\%. The median values of stellar masses of
volume-limited samples of LSBGs and HSBGs are both 2.19$\times$10$^{10}$
$M_\odot$. This result in our volume-limited samples is different
from Liang et al. (2010) who showed that galaxies with lower surface
brightness have smaller stellar masses generally. This difference
may be because they use magnitude-limited samples that could be affected
by the completeness, while we use volume-limited samples to avoid
such selection effects. Moreover, the nearly same stellar masses distributions
between LSBGs and HSBGs can help to get a more fair comparison in
the following sections since the affecting of stellar masses is nearly
removed.

The stellar metallicities [log($Z/Z_\odot$)] of LSBGs also span nearly the same
range as that of HSBGs (Fig. \ref{fig.his}b), however, the median value
of stellar metallicities for LSBGs is -1.95 which is 0.11 dex (1.3 times)
lower than that of HSBGs (-1.84). Kolmogorov-Smirnov tests show that the
different mean of distributions between LSBGs and HSBGs is 16.0\%. The
result that LSBGs have lower stellar metallicities is consistent with the
lower metallicities of LSBGs found in gas-phase (McGaugh 1994, Galaz et
al. 2006, Haberzettl et al. 2007b, Liang et al. 2010). Liang et al. (2010)
found that the decreasing of metallicities with decreasing surface brightness
could be due to the decreasing stellar masses with decreasing surface brightness.
In our volume-limited samples, however, LSBGs have the similar stellar masses
distributions with HSBGs, yet they still have lower metallicities. Thus,
lower stellar masses is not enough to explain the lower metallicities of
LSBGs. The LSBGs could have different star formation histories from HSBGs.

Fig. \ref{fig.his}c shows that the LSBGs have lower effective dust
attenuations in $z$ band,. The median value is 0.34 mag, which is
0.11 mag smaller than that of HSBGs whose median value of effective
dust attenuations in $z$ band is 0.45. Kolmogorov-Smirnov tests
show that the different mean of distributions between LSBGs and HSBGs
is up to 26.4\%.  The lower effective dust attenuations in red
band ($z$ band) is consistent with the results in $V$ band (e.g. Liang
et al. 2010), which means that LSBGs could contain less dust than HSBGs 
(McGaugh 1994; Hinz et al. 2007).

In Fig. \ref{fig.his}d, we show the distributions of dust-corrected $z$ 
band mass-to-light ratios, shadowed and white regions are for LSBGs and HSBGs
respectively. The LSBGs tend to have higher mass-to-light ratios than HSBGs.
Kolmogorov-Smirnov tests show that the different mean of distributions
between LSBGs and HSBGs is 23.7\%. The median value of stellar mass to
$z$ band luminosity ratios ($M_*/L_z$) is 1.11 that is 0.15 (1.4 times)
higher than that of HSBGs with median $M_*/L_z$ of 0.96. The high $M_*/L_z$
ratios either point toward galaxies early in their evolution that have not
converted gas into stars or reflect more recent accumulation of gas by
an evolved stellar population (Burkholder et al. 2001). This result may 
also indicate different star formation histories between LSBGs and HSBGs.  

\subsection{Gas contents}

\begin{figure*}
\centering
\includegraphics[width=6.0cm]{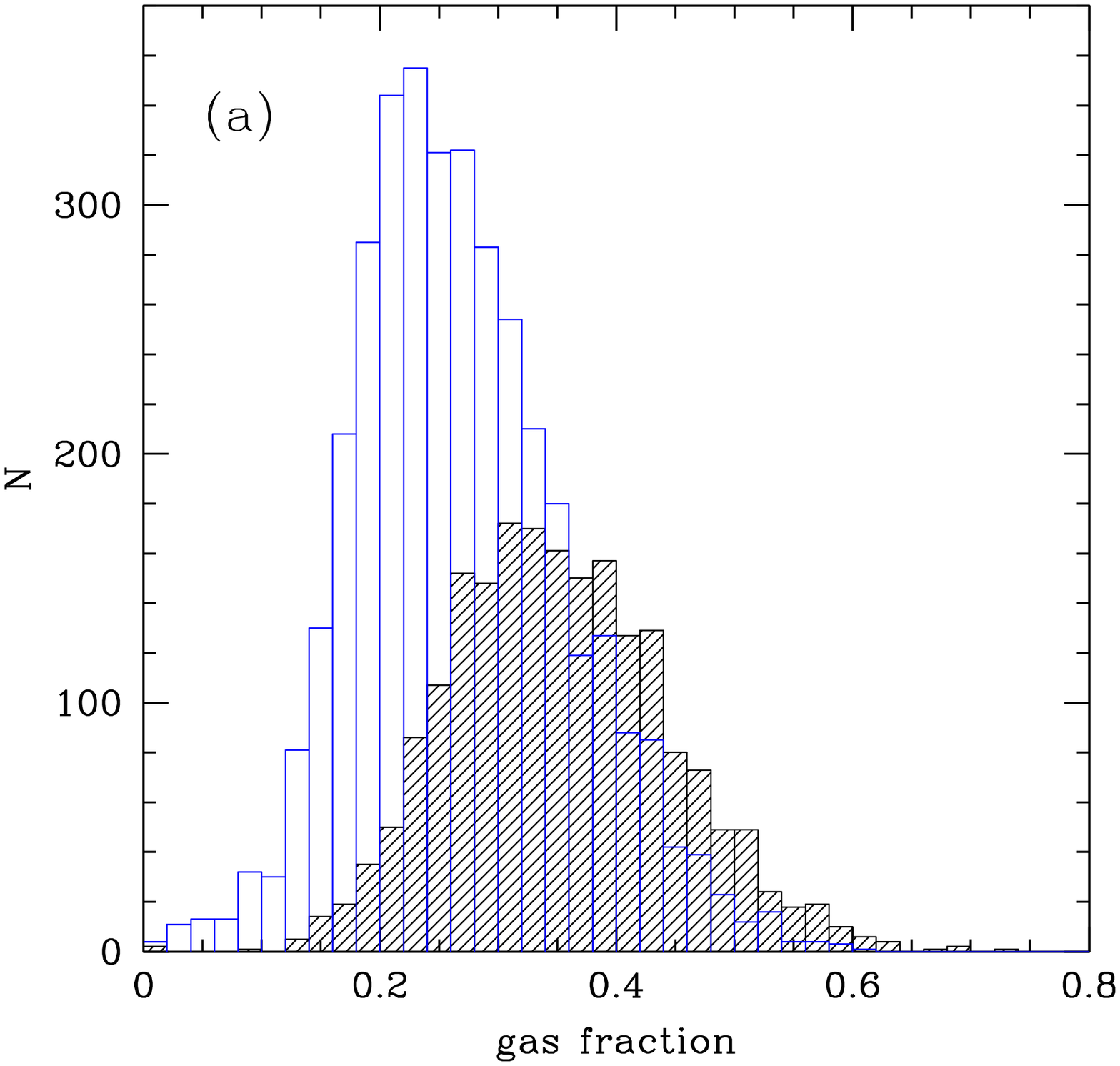}
\includegraphics[width=7.0cm, height=6.0cm]{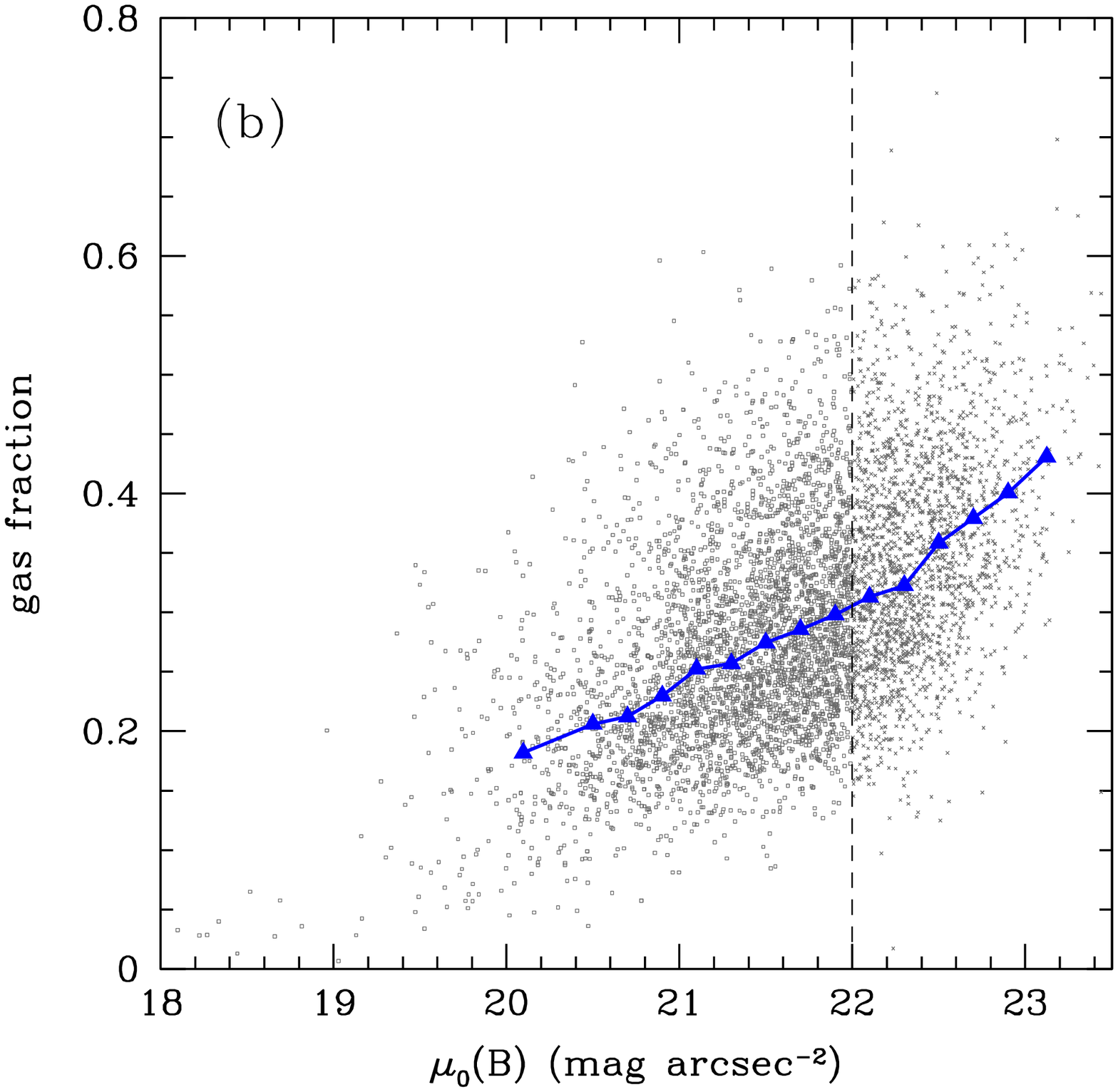}
\caption{The histogram distributions of gas fraction (a), shadowed and 
white regions are for LSBGs and HSBGs respectively. The relation between gas
fraction and surface brightness (b). The solid triangles and line denote 
the median values with surface brightness bins. The vertical dashed line marks
the range of LSBGs and HSBGs.}
\label{fig.gasf}
\end{figure*}
Gas fraction can be used to quantify galaxy evolution (e.g. McGaugh \& de Blok 
1997; Burkholder et al. 2001). Assuming a constant $M_*/L$, it is said that LSBGs 
are considered as gas-rich galaxies, which may result from the fact that LSBGs 
have not converted gas into stars effectively. In our work, we first calculate 
the HI gas-to-stellar mass ratio (log[$G/S$]) by using the formula of Zhang et al. 
(2009, see their Eq. 4 for more details)
\begin{equation}
\label{eq.gs} 
\log (G/S)= -1.73238(g - r) + 0.215182 \mu_i - 4.08451,
\end{equation}
where $G$ and $S$ refer to HI mass ($M_{HI}$) and stellar mass ($M_*$),
and $g$, $r$ are the apparent magnitudes from SDSS in $g$, $r$ band,
both of which are $K$-corrected and Galactic reddening corrected (Blanton 
et al. 2005). The $\mu_i$ is surface brightness in SDSS $i$ band, which 
is defined as $\mu_i = m_i + 2.5 \log (2\pi R_{50}^2)$, where $m_i$ is 
the apparent magnitude in $i$-band that is also $K$-corrected and Galactic 
reddening corrected (Blanton et al. 2005), and $R_{50}$ is the radius 
(in units of arcsec) enclosing 50 per cent of the total Petrosian $i$-band flux. 
Then the HI mass can be calculated from log($G/S$) because the stellar mass is 
available (Kauffmann et al. 2003). At last, one can get the gas fraction 
from $f_g = M_g/(M_g+M_*)$, where $M_g = $ 1.4$M_{HI}$ (e.g. McGaugh \& 
de Blok 1997; Schombert et al. 2001).

We show the histogram of the calculated gas fraction in Fig. \ref{fig.gasf}a, 
from which we can see that the gas fraction of LSBGs span nearly the same range 
as that of HSBGs but more likely to distribute in the higher gas fraction region.
We try our best to find related work on gas fraction of LSBGs to compare with our results, 
and found three much related ones (i.e. McGaugh \& de Blok 1997; Schombert et al. 
2001; Burkholder et al. 2001). The range of gas fraction of our volume-limited 
LSBGs is approximate from 15\% to 70\%, which is very similar to the results 
in McGaugh \& de Blok (1997) for small sample of LSBGs (17\% to 77\%). Schombert 
et al. (2001) found that a majority of the galaxies in their disk samples have 
gas fraction below 50\% peaked at 30\%, whereas in the LSB dwarf galaxies over 
90\% of the galaxies have gas fraction greater than 30\%. For our volume-limited 
LSBGs, most have gas fraction below 70\% with a median value of 34.6\%. This 
result is higher than that of the disk sample of Schombert et al. (2001), but 
lower than that of their LSB dwarf galaxies. The reason is that our volume-limited 
LSBGs contain more brighter galaxies without any dwarf galaxies (Zhong et al. 2008). 
The median value of gas fraction of our LSBGs is quite similar to that of 
McGaugh \& de Blok (1997) with small sample of LSBGs (median value of 40.1\%), 
only 5.5\% higher than ours (34.6\%). Moreover, the median value of gas fraction 
for the our LSBGs is 8.7\% higher than that of HSBGs (median value of 25.9\%), 
thus, LSBGs are indeed gas-rich galaxies. Kolmogorov-Smirnov tests show that 
the different mean of distributions between LSBGs and HSBGs is bigger, up to 48.7\%. 

Furthermore, there is a tight correlation between gas fraction and surface brightness 
(the Spearman rank order correlation coefficient is 0.51) that galaxies with 
lower surface brightness have higher gas fraction (Fig. \ref{fig.gasf}b). This
relation has been found in some previous studies with very small samples. For
example, McGaugh \& de Blok (1997) found the correlation coefficient between
gas fraction and surface brightness is 0.63, which is a little tighter than ours.
Schombert et al. (2001) used chemical and spectrophotometric models from Boissier
\& Prantzos (2000) also predict the general trend of higher gas fraction with
fainter surface brightness.

Though the correlation we find shows more scatter than those found by McGaugh \& 
de Blok (1997), Schombert et al. (2001), and Burkholder et al. (2001), our conclusion 
is similar to theirs: LSBGs tend to be gas-rich. The high gas fraction of LSBGs could 
indicate that either these galaxies have experienced delays in formation and are 
just beginning to form stars or their ongoing star formation is inefficient and/or 
sporadic (Burkholder et al. 2001).

Moreover, the slope of gas fraction vs. surface brightness for LSBGs is a little 
steeper than that of HSBGs. 

\subsection{Star formation rates}
\label{sfr}

\begin{figure*}
\centering
\includegraphics[width=6.0cm]{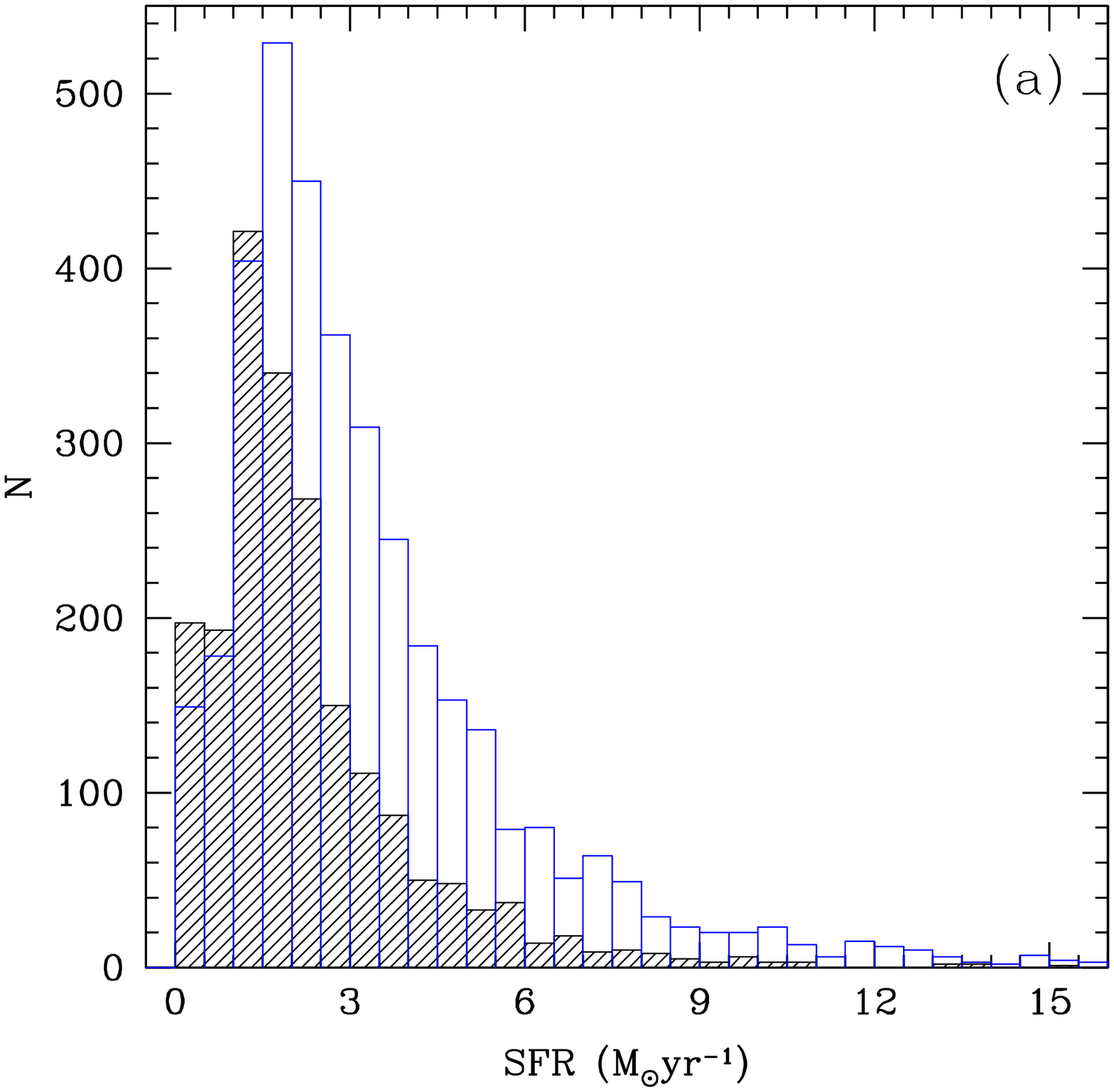}
\includegraphics[width=7.0cm, height=6.0cm]{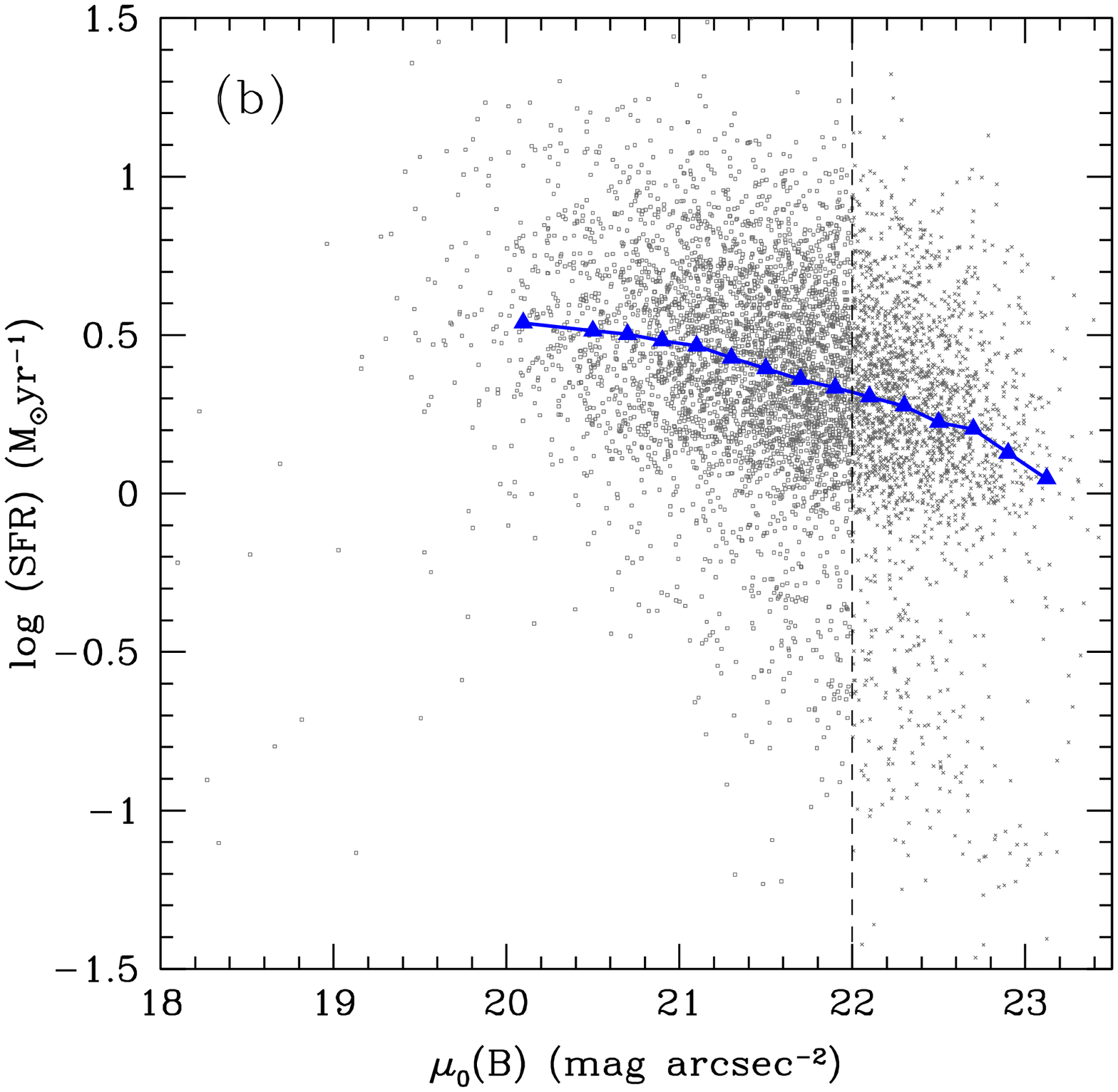}
\caption{The histogram distributions of SFRs (a), shadowed and
white regions are for LSBGs and HSBGs respectively. The relation
between SFRs and surface brightness (b). The open squares, crosses, solid
triangles, solid line, and dashed line are the same as Fig. \ref{fig.gasf}.}
\label{fig.SFRs}
\end{figure*}

It is showed that the typical gas surface densities for LSBGs are
below the Kennicutt criterion for ongoing star formation (Kennicutt
1989; van der Hulst et al. 1993) which results in a suppressed
current SFRs (Bossier et al. 2008; Schombert et al. 2011). Thus the
SFRs of LSBGs should be lower than that of HSBGs. In Fig.
\ref{fig.SFRs}a, we show the histogram distributions of SFRs for
both LSBGs (shadowed regions) and HSBGs (white regions). We can see
that LSBGs are more likely to have lower SFRs. Kolmogorov-Smirnov
tests show that the different mean of distributions between LSBGs
and HSBGs is 30.7\%. The median value of SFRs for LSBGs is 1.77
$M_\odot$ yr$^{-1}$ which is 0.86 $M_\odot$ yr$^{-1}$ smaller than
that of HSBGs that have median SFRs of 2.63 $M_\odot$ yr$^{-1}$.
Moreover, the SFRs drop quickly with the decreasing surface brightness
(see Fig.~\ref{fig.SFRs}b), the Spearman rank order correlation
coefficient is -0.32, which could suggest that higher surface
brightness galaxies are undergoing stronger star forming activities
than the lower ones.

 However the SFRs of LSBGs span a wide range, and the SFRs of LSBGs are
not as low as the results in the previous studies (e.g. 0.02-0.8 $M_\odot$
yr$^{-1}$, van den Hoek 2000). The reason of the relatively higher
SFRs for LSBGs may be that we use volume-limited sample of LSBGs,
which select the brighter LSBGs.

\subsection{Specific SFRs}

\begin{figure*}
\centering
\includegraphics[width=6.0cm]{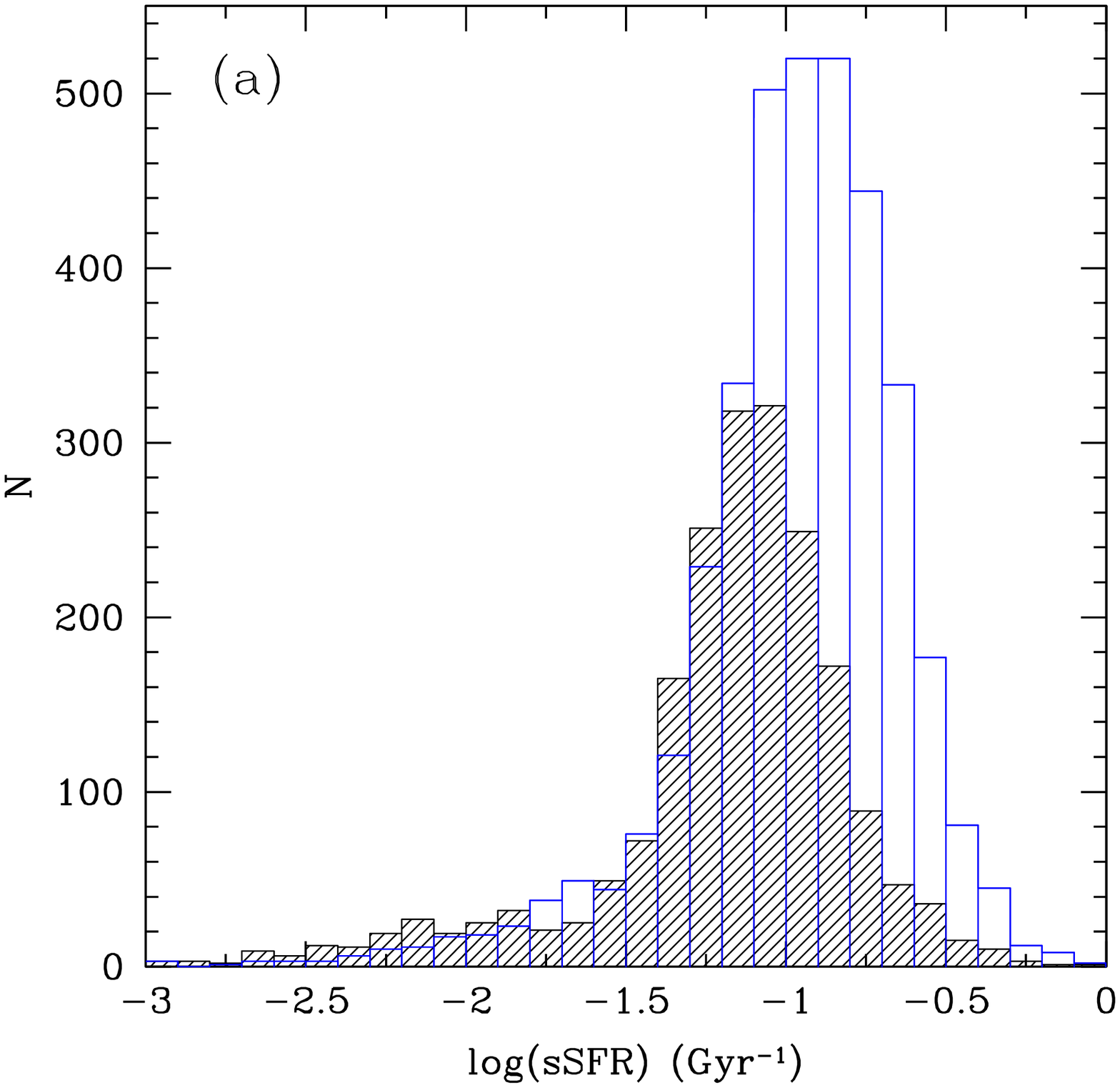}
\includegraphics[width=7.0cm, height=6.0cm]{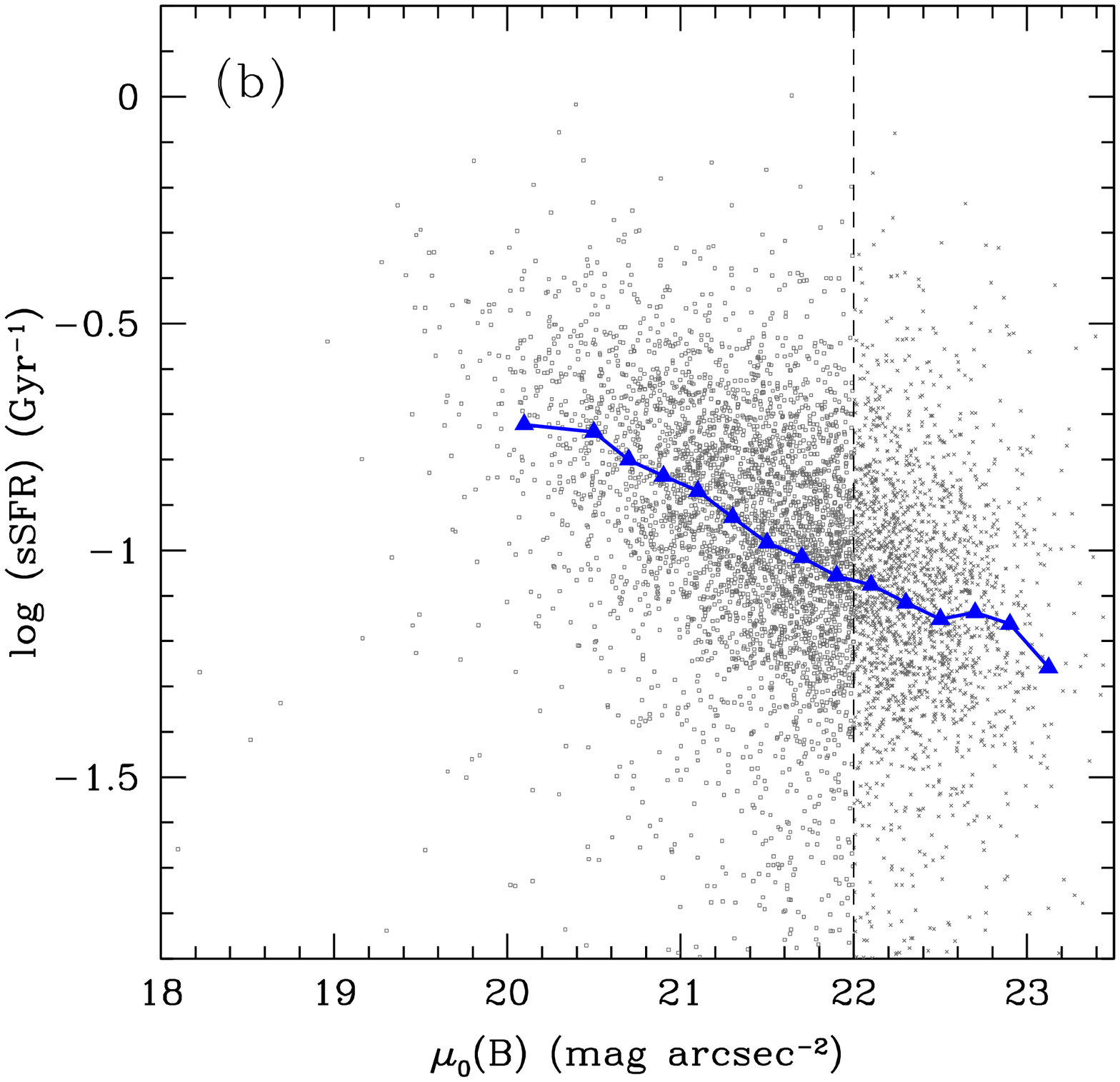}
\caption{The histogram distributions of specific SFRs (a),
shadowed and white regions are for LSBGs and HSBGs respectively.
The relation between specific SFRs and surface brightness (b). The open
squares, crosses, solid triangles, solid line, and dashed line
are the same as Fig. \ref{fig.gasf}.}
\label{fig.sSFRs}
\end{figure*}

Given the strong correlation between SFRs and stellar masses, it is clear that
by normalizing the SFRs by the stellar masses, one can more easily study
the relationship between star formation activity and the physical parameters
of the galaxies (Brinchmann et al. 2004). Although the distributions of
stellar masses for LSBGs and HSBGs in the volume-limited samples are similar,
we could further remove the effect of stellar masses by comparing the specific
SFRs of LSBGs with that of HSBGs.

The specific SFRs are also from Brinchmann et al. (2004). The median value
of specific SFRs for LSBGs (shadowed regions) is -1.12 Gyr$^{-1}$ (Fig.
\ref{fig.sSFRs}a) which is 0.18 Gyr$^{-1}$ smaller than that of HSBGs
(white regions). Kolmogorov-Smirnov tests show that the different mean of
distributions between LSBGs and HSBGs is up to 34.9\%. This result is
consistence with the result of Sect \ref{sfr}, which shows that the LSBGs
have smaller SFRs than HSBGs but nearly the same stellar masses as HSBGs.
Hence LSBGs should have lower specific SFRs than that of HSBGs. Furthermore,
the specific SFRs decrease rapidly with the decreasing of surface brightness,
i.e. the lower the surface brightness are, the lower specific SFRs they have
(Fig. \ref{fig.sSFRs}b). LSBGs with lower specific SFRs could be also related
to Kennicutt criterion for ongoing star formation, and LSBGs form stars in
longer periods of time (e.g. Galaz et al. 2011). The slope of specific SFRs
vs. surface brightness is a little steeper than the slope of SFRs vs. surface
brightness, the Spearman rank order correlation coefficient is -0.40. This may
be due to the effects of stellar masses that have been further removed in
calculating specific SFRs. Both Fig.~\ref{fig.SFRs} and Fig. \ref{fig.sSFRs}
show that the current star forming activities of HSBGs are more active
than that of LSBGs.

The specific SFRs has often been rephrased in terms of the present
to past-average star formation rate, which immediately gives an indication
of the past star formation history of the galaxy and its relation to present-day
activity, e.g. the birthrate parameter $b$ (the present-to-past average SFR
ratio, Kennicutt et al. 1994; Brinchmann et al. 2004; Galaz et al. 2011). 
Following the Eq. 2 of Galaz et al. (2011), we found that there is a fraction
of 88.2\% for LSBGs with $b$ less than 1, while the corresponding fraction of
HSBGs is 66.8\%. It means that the average $b$ parameter is higher in 
HSBGs than in LSBGs (Galaz et al. 2011), and the average value of $b$ parameter 
for our LSBGs is similar to that of Galaz et al. (2011), 0.59 vs. 0.65.

\section{Discussions}
\label{discussions}

The previous studies about stellar metallicities, $M_*/L$, SFRs and gas fractions
of LSBGs have been traditionally carried out with very small samples. With the
advent of the large sky survey of the SDSS, it is now possible to dramatically
extend these studies in size, and it may also improve the qualities because their
large amount of high-quality data will be undoubtedly important to allow 
to study the properties of those galaxies more carefully.

\begin{figure}
\centering
\includegraphics[width=7.0cm]{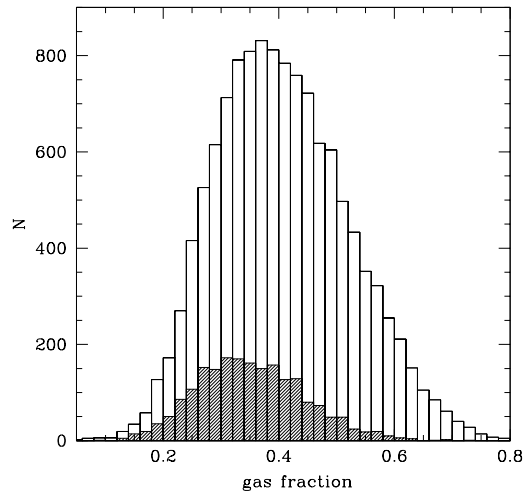}
\caption{The histogram distributions of gas fraction for
volume-limited LSBGs (2,021, shadowed regions) and all LSBGs (12,282, 
white regions).}
\label{fig.com}
\end{figure}

Moreover nearly all the previous studies of stellar metallicities, $M_*/L$, SFRs
and gas fractions for LSBGs are based on magnitude-limited sample, which could
be affected by selection effects. We try to avoid the selection
effects on the studies of LSBGs by using volume-limited samples in this work. And we extend 
much more the number of LSBGs, which help well to study the statistical properties
of LSBGs. Fortunately, our volume-limited samples of LSBGs and HSBGs have very
similar stellar masses that can help to get a more fair comparison between their other
properties because it nearly removes the effect of stellar masses.

However, could the different properties between LSBGs and HSBGs studied above
be caused by the fact that the volume-limited sample contains a larger fraction
of gas-poor galaxies? Thus, we calculate the gas fraction for all the parent LSBGs 
(i.e 12,282 LSBGs) and HSBGs (i.e. 18,051 HSBGs) using the same formula 
of Zhang et al. (2009). In Fig. \ref{fig.com}, as an example, we show the 
histogram distributions of gas fraction for volume-limited LSBGs 
(2,021, shadowed region) and all parent LSBGs (12,282, white region). The median 
value of gas fraction for all parent LSBGs is 39.8\%, which is only 5.2\% higher 
than that of volume-limited LSBGs (34.6\%). And the median value of gas fraction for 
all HSBGs is 29.0\% that is only 3.1\% higher than that of volume-limited HSBGs (25.9\%). 
Therefore, although we could lose some gas-rich LSBGs and HSBGs when selecting 
volume-limited samples, this would not affect our results too much because we 
just lose a very small fraction of gas-rich galaxies, and the fraction we lose 
for LSBGs and HSBGs is very similar. 

Furthermore, we compare the gas fraction in our volume-limited LSBGs with that of 
small sample of LSBGs in McGaugh \& de Blok (1997) to see whether our volume-limited 
LSBGs have quite different gas fraction from the previous studies or not. For example, 
the gas fraction in McGaugh \& de Blok (1997) for LSBGs is from 17\% to 77\% which 
is very similar to the range of our volume-limited LSBGs (15\% to 70\%). And the 
median value of gas fraction for the LSBGs in McGaugh \& de Blok (1997) is 40.1\%, 
which is not very different from our volume-limited LSBGs (34.6\%), only 5.5\% higher 
than ours. This means that our volume-limited LSBGs could be a reasonable sample to 
show the properties of LSBGs.

\section{Summaries}
\label{conclusion}

In this paper, we continue our studies on the properties of a large
sample of LSBGs from SDSS. We select a relatively large volume-limited
complete sample of LSBGs from SDSS-DR4 following Zhong et al. (2008)
to study their properties and compare their properties with HSBGs.
 This large sample of LSBGs is useful to study the statistical properties
of LSBGs. And by using the volume-limited samples of LSBGs and HSBGs,
we can avoid the bias introduced from the differences in the distributions
of the redshift and absolute magnitude for our samples. Moreover our volume-limited
LSBGs have similar stellar masses to HSBGs that further remove the effect
of stellar mass. The results can be summarized as follows.

\begin{enumerate}

\item LSBGs tend to have lower effective dust attenuations in $z$ band with
median value of 0.34 mag, which is 0.11 lower than that of HSBGs. This means 
that LSBGs contain less dust than HSBGs.

\item The distributions of stellar masses are nearly the same for the two 
samples, both of which are with median values of 2.19 $\times$ 10$^{10}$ $M_\odot$. 
However, the median value of stellar metallicities, log($Z/Z_\odot$), for 
LSBGs is still 0.11 dex lower than that of HSBGs. And LSBGs have higher 
mass-to-light ratios (in $z$ band, Kauffmann et al. 2003). The median value 
of mass-to-light ratios in $z$ band ($M_*/L_z$) after extinction correction 
is 1.11, which is 0.15 higher than that of HSBGs. 

\item  LSBGs are likely to have higher gas fraction than that of HSBGs,
with the median value of gas fraction 8.7\% higher. And there is a tight correlation
between gas fraction and surface brightness that galaxies with lower surface
brightness have higher gas fraction, and the slope of gas fraction vs. surface
brightness for LSBGs is slightly steeper than that of HSBGs. Although
we may lose some gas-rich LSBGs when selecting our volume-limited LSBGs,
however, we find that the gas-fraction of our volume-limited LSBGs is
not much different from all the parent LSBGs (34.6\% vs. 39.8\%). Moreover our large sample
of LSBGs have the similar gas fraction to the McGaugh \& de Blok (1997) with 
a small sample of LSBGs.

\item LSBGs have lower SFRs with median value of 1.77 $M_\odot$ yr$^{-1}$, 
which is 0.86 $M_\odot$ yr$^{-1}$ lower than that of HSBGs which suggest 
that HSBGs underwent more recent star forming activities than LSBGs.
However the median value of SFRs for LSBGs in our work is not as low
as in previous studies. Moreover, the SFRs decrease with decreasing
surface brightness, i.e. the higher the surface brightness are, the
stronger present star forming activities they have.

\item The specific SFRs of LSBGs are also lower than that of HSBGs, and the 
specific SFRs also decrease with decreasing surface brightness, but the 
slope of the relation between specific SFRs and surface brightness is a little steeper than 
that of SFRs and surface brightness, which may be due to the effects of 
stellar masses that are further removed in calculating specific SFRs.

\end{enumerate}

In summary, LSBGs have different star formation histories from HSBGs
and HSBGs may have stronger star forming activities than LSBGs.

%----------------------END-Conclusion-----------------------------------------%

\section*{Acknowledgements}
We are grateful to the referee for constructive comments from which this
paper has benefitted. This work was supported by the National Natural Science 
Foundation of China (Grant Nos. 10933001, 10973006, 10973015 and 10673002); the
National Basic Research Program of China (973 Program; Nos. 2007CB815404
and 2007CB815406); and the Young Researcher Grant of National Astronomical
Observatories, Chinese Academy of Sciences. We thank the useful MPA/JHU
catalogs.

Funding for the SDSS and SDSS-II has been provided by the Alfred P.
Sloan Foundation, the Participating Institutions, the National Science
Foundation, the U.S. Department of Energy, the National Aeronautics
and Space Administration, the Japanese Monbukagakusho, the Max Planck
Society, and the Higher Education Funding Council for England. The
SDSS Web site is http://www.sdss.org/. The SDSS is managed by the
Astrophysical Research Consortium for the Participating Institutions.
The Participating Institutions are the American Museum of Natural
History, Astrophysical Institute Potsdam, University of Basel,
University of Cambridge, Case Western Reserve University, University
of Chicago, Drexel University, Fermilab, the Institute for Advanced
Study, the Japan Participation Group, Johns Hopkins University,
the Joint Institute for Nuclear Astrophysics, the Kavli Institute
for Particle Astrophysics and Cosmology, the Korean Scientist Group,
the Chinese Academy of Sciences (LAMOST), Los Alamos National Laboratory,
the Max-Planck-Institute for Astronomy (MPIA), the Max-Planck-Institute
for Astrophysics (MPA), New Mexico State University, Ohio State University,
University of Pittsburgh, University of Portsmouth, Princeton University,
the United States Naval Observatory, and the University of Washington.

%\section*{References}

% REFERENCES -----------------------------------------------------------------%
\footnotesize
\bibliographystyle{elsarticle-harv}
%\bibliographystyle{elsarticle-num-names}
%\bibliography{/Users/phall/OLDWORK/AASTEX/pathall}
%\bibliography{/home/phall/OLDWORK/AASTEX/pathall}
%\bibliography{/home/jar/project1/pathall}

\label{lastpage}

\end{document}